\documentclass[aps,superscriptaddress,eqsecnum, twocolumn]{revtex4-1}
\usepackage{graphicx}
\usepackage{stmaryrd}
\usepackage{amsmath, amsfonts}
\usepackage{mathrsfs}
\usepackage{braket}
\usepackage{color}
\usepackage{xspace}
\usepackage{amscd}

\usepackage{bm}
\definecolor{lightblue}{rgb}{0.13, 0.26, 0.99}

\usepackage[
colorlinks=true,
urlcolor=blue,
citecolor=blue,
linkcolor=blue,
hyperfootnotes=false]{hyperref}

\newcommand{\im}{\operatorname{Im}}

\allowdisplaybreaks

\begin{document}

\title{Polarization amplitude near quantum critical points}

\author{Shunsuke C. Furuya}
\affiliation{Condensed Matter Theory Laboratory, RIKEN, Wako, Saitama 351-0198, Japan}
\author{Masaaki Nakamura}
\affiliation{Department of Physics, Ehime University Bunkyo-cho 2-5, Matsuyama, Ehime 790-8577, Japan}
\date{\today}
\begin{abstract}
    We discuss the polarization amplitude of quantum spin systems in one dimension.
    In particular, we closely investigate it in gapless phases of those systems based on the two-dimensional conformal field theory.
    The polarization amplitude is defined as the ground-state average of a twist operator which induces a large gauge transformation attaching the unit amount of the $U(1)$ flux to the system.
    We show that the polarization amplitude under the periodic boundary condition is sensitive to perturbations around the fixed point of the renormalization-group flow rather than the fixed point itself even when the perturbation is irrelevant.
    This dependence is encoded into the scaling law with respect to the system size.
    In this paper, we show how and why the scaling law of the polarization amplitude encodes the information of the renormalization-group flow.
    In addition, we show that the polarization amplitude under the antiperiodic boundary condition is determined fully by the fixed point in contrast to that under the periodic one and that it visualizes clearly the nontriviality of spin systems in the sense of the Lieb-Schultz-Mattis theorem.
\end{abstract}
\maketitle

\section{Introduction}\label{sec:intro}

\begin{table*}[t!]
    \centering
    \begin{tabular}{cccccc} \hline \hline
       `lattice model' & `$\mathcal H'$ at $L\to+\infty$' & `$\mathcal H'$ in RG' & `interaction range' & `$q$' & `power $\beta_q$' \\ \hline
         XXZ chain & present & irrelevant  & local & 2 & $4K-2$ \\
         fine-tuned $J_1$-$J_2$ XXZ chain & absent & irrelevant & local & 2 & $4K$ \\
         Haldane-Shastry & absent & irrelevant & nonlocal & 2 & $4K-1$ \\
         \hline
         ladder on QCP & absent & relevant & local & $1$ & $2K$ \\
         ladder off QCP & present & relevant & local & $1$  & $2K-2$ ($<0$) \\
         \hline \hline
    \end{tabular}
    \caption{Comparison of the scaling law $z^{(q)} \propto  (2\pi/L)^{\beta_q}$ in four spin-chain models: the XXZ spin chain, the $J_1$-$J_2$ XXZ spin chain, the Haldane-Shastry model, and the spin ladder on and off the quantum critical point (QCP). While all these models have the same fixed-point theory, they disagree in the scaling law. The second column refers to the existence of the most relevant perturbation \eqref{H'_chain_comp} in the first three models and $\mathcal H' \propto \int dx \cos(2\sqrt 2\phi_+)$ in the last two. 
    The third columns refers to the relevance of $\mathcal H'$ in the RG sense and the fifth column to the smallest positive $q$ that makes $z^{(q)}$ nonzero under the periodic boundary condition.
    The rightest column shows the power $\beta_q$ for that $q$.
    }
    \label{tab:scaling}
\end{table*}

The electric polarization is one of the most fundamental quantities in condensed-matter systems.
It is sensitive to the conductivity of electrons and thus allows us to diagnose whether the system is in a conducting phase or in an insulating phase~\cite{resta, resta-sorella}.
The polarization is regarded as a bulk property of the system~\cite{watanabe_polarization} and therefore it should be well-defined under the periodic boundary condition.
However, the naive definition of the polarization as the spatial integral of the product of the position $x$ and the charge density $n(x)$ suffers from an ambiguity of the position operator under the periodic boundary condition.
It was thus proposed~\cite{resta,resta-sorella} to define the polarization through the polarization amplitude $z$ given by
\begin{align}
    z &:=\braket{\psi_0|e^{i\frac{2\pi}{L} \mathcal P}|\psi_0},
    \label{def_z}
\end{align}
where $\ket{\psi_0}$ is the ground state of the system, $L$ is the length of the system along an arbitrarily chosen direction, and $\mathcal P$ is the  polarization operator along that direction, $\mathcal P = \int_0^L dx \, x n(x)$.
Using the polarization amplitude, one may write the polarization as $(L/2\pi)\im \ln z$.
The polarization amplitude has attracted much attention of theoretical physicists~\cite{aligia_polarization_amp, oshikawa_lsm_flux, kobayashi_polarization, Dora_polarization_amp, yahayavi_polarization, hetenyi_polarization_amp}

The polarization amplitude also plays an important role in magnetic insulators.
Dealing with magnetic excitations as charged particles,
we can adapt the polarization amplitude to distinguish gapless and gapped ground states of quantum spin systems.
Moreover, a generalized polarization amplitude,
\begin{align}
    z^{(q)} &:= \braket{\psi_0|U^q|\psi_0},
    \label{def_zq}
\end{align}
with an integer $q$ and
\begin{align}
    U^q &:=\exp\biggl(i\frac{2\pi q}{N} \sum_{j=1}^{N} j S_j^z \biggr),
    \label{def_Uq}
\end{align}
is useful to distinguish symmetry-protected topological phases and topologically trivial gapped phases~\cite{nakamura-voit, nakamura-todo, tasaki_twist}.
In addition, one can expect that $z^{(q)}$ is also related to the ``gappability'' of the ground state, that is, the possibility of having a unique symmetric gapped ground state.
This notion is closely related to the Lieb-Schultz-Mattis (LSM) theorem~\cite{lsm,oshikawa_lsm_flux, hastings_lsm} and also the symmetry protection of gapless phases~\cite{furuya_anomaly}.
In fact, the operator $U$ played a key role in the original proof of the LSM theorem~\cite{lsm}.

The growing awareness of the LSM theorem and related phenomena in condensed-matter physics~\cite{ cho_lsm, metlitski_lsm, yao_anomaly, tanizaki_su3, kobayashi_lsm} leads theoretical physicists naturally to studies on the polarization amplitude \eqref{def_zq} in gapless quantum many-body systems~\cite{kobayashi_polarization, Dora_polarization_amp,hetenyi_polarization_amp, hetenyi_polarization_amp}.
Recently, Kobayashi \textit{et al.}~\cite{kobayashi_polarization} investigated the polarization amplitude \eqref{def_zq} in gapless phases of one-dimensional quantum spin systems with a finite chain length $L$ on the basis of equivalence of quantum spin chains and interacting spinless fermion systems in one dimension~\cite{Giamarchi_book}.
They found an interesting scaling law of the polarization amplitude with respect to $L$:
\begin{align}
    z^{(q)} \propto \biggl(\frac{2\pi}{L}\biggr)^{\beta_q},
    \label{zq_scaling}
\end{align}
with a power $\beta_q$.
It was found that
the scaling law in  $S=1/2$ XXZ spin chain differs from that in an $S=1/2$ $J_1$--$J_2$ XXZ chain although the ground states of those systems belong to the same quantum phase described by a low-energy effective field theory known as the Tomonaga-Luttinger (TL) liquid~\cite{Giamarchi_book}.
This observation implies that the scaling law of the polarization amplitude is sensitive to some high-energy degrees of freedom which can usually be ignored without any problem in the low-energy effective theory.
Currently, the scaling law of the polarization amplitude in gapless phases is yet to be understood despite its potential importance in the recent development around the LSM theorem.

Also, the authors reported quite recently in Ref.~\cite{nakamura_jump} that the polarization amplitude exhibits a discontinuity by a universal amount at a quantum critical point described by the TL liquid.
Such a universal jump shows up when the antiperiodic boundary condition is imposed~\cite{kitazawa_tbc}, that is, when a $1/2$-unit U(1) flux is attached to the system.
While the universal jump was investigated analytically and numerically,
the low-energy field-theoretical description of the universal jump was not discussed in Ref.~\cite{nakamura_jump}.

In this paper, considering those recent developments in theoretical studies on the polarization amplitude, we develop a theory of the polarization amplitude in and near quantum critical phases of quantum spin systems in one dimension.
Our discussion is based on the two-dimensional conformal field theory (CFT).
We deal with four specific models: the $S=1/2$ XXZ spin chain, the $S=1/2$ $J_1$--$J_2$ spin chain, the Haldane-Shastry model~\cite{haldane_HS, shastry_HS}, and an $S=1/2$ spin ladder.
We show that the scaling law is universal but sensitive to perturbations around the fixed point of the renormalization-group (RG) flow.
The scaling law in those systems is summarized in table~\ref{tab:scaling}.
In addition, we point out another interesting aspect of the polarization amplitude.
The polarization amplitude under the antiperiodic boundary condition is determined by the fixed-point theory alone, which is in sharp contrast to the case of the periodic boundary condition.

This paper is organized as follows.
In Sec.~\ref{sec:XXZ}, we deal with the simplest case of the $S=1/2$ XXZ spin chain.
We discuss a simple perturbative approach to the scaling law consistent with the numerical results of Ref.~\cite{kobayashi_polarization}.
To get further insight into the scaling law, we consider another model, the $S=1/2$ $J_1$-$J_2$ XXZ chain, where the perturbation considered in Sec.~\ref{sec:XXZ} is eliminated in the $L\to + \infty$ limit by a fine tuning of parameters.
We show that the fine tuning modifies the scaling law and clarifies the mechanism of modification in Sec.~\ref{sec:J1J2}.
In Sec.~\ref{sec:HS}, we focus on a specific model known as the Haldane-Shastry model which is free from any perturbation and realizes the fixed-point theory.
The results in Secs.~\ref{sec:XXZ}, \ref{sec:J1J2}, and \ref{sec:HS} are adapted to a spin ladder in Sec.~\ref{sec:ladder}.
In Sec.~\ref{sec:APBC}, we discuss another property of the polarization amplitude independent of perturbations to the fixed-point theory.
Finally, we summarize this paper in Sec.~\ref{sec:summary}.

\section{$S=1/2$ XXZ chain}\label{sec:XXZ}

Let us consider the $S=1/2$ XXZ chain described by the Hamiltonian,
\begin{align}
    \mathcal H_{\rm XXZ} &= J \sum_{j=1}^N (S_j^x S_{j+1}^x + S_j^y S_{j+1}^y + \Delta S_j^z S_{j+1}^z),
    \label{H_XXZ}
\end{align}
with $J>0$ and $-1<\Delta \le 1$.
Here, $\bm S_j$ is the spin defined on the $j$th site of the spin chain with the finite chain length $L=Na_0$ ($a_0$ is the lattice spacing).
Hereafter, we employ the unit $a_0=\hbar =1$ for notational simplicity but we will restore them if needed for logical clarity.

The $S=1/2$ XXZ chain \eqref{H_XXZ} is well known to have the gapless ground state known as the TL liquid.
The $S=1/2$ XXZ chain \eqref{H_XXZ} provides us the simplest example to discuss the scaling law in interacting systems.
Note that we impose the periodic boundary condition,
\begin{align}
    \bm S_{L+1} = \bm S_1.
    \label{pbc_spin}
\end{align}

The TL liquid state is described by the following Hamiltonian,
\begin{align}
    \mathcal H_{\rm TL}^\ast &= \frac{v}{2\pi} \int_0^L dx \, \biggl(K(\partial_x \theta)^2 + \frac 1K(\partial_x \phi)^2 
    \biggr).
    \label{H_XXZ_fixedpoint}
\end{align}
The fields $\theta(x)$ and $\phi(x)$ satisfy the commutation relation,
\begin{align}
    [\phi(x), \partial_y\theta(y)] = i\pi \delta(x-y),
    \label{comm_phi}
\end{align}
and are related to the spin operator through~\cite{Giamarchi_book}
\begin{align}
    S_j^z &= \frac{a_0}{\pi}\partial_x \phi + (-1)^j a_1 \cos(2\phi),
    \label{Sz2phi} \\
    S_j^\pm &= e^{\pm i\theta} \bigl[ (-1)^j b_0 + b_1 \cos(2\phi)
    \bigr].
    \label{Spm2theta}
\end{align}
$a_1$, $b_0$, and $b_1$ are nonuniversal and numerically determined~\cite{hikihara_coeff1, takayoshi_coeff, hikihara_coeff2}.
The periodic boundary condition \eqref{pbc_spin} turns into
\begin{align}
    \phi(x+L) &= \phi(x) + \pi N_\phi,
    \label{pbc_phi} \\
    \theta(x+L) &= \theta(x) + 2\pi N_\theta,
    \label{pbc_theta}
\end{align}
with $N_\phi, N_\theta \in \mathbb Z$.
The velocity $v$ and the TL parameter $K$ are exactly obtained in the $S=1/2$ XXZ chain~\cite{Giamarchi_book}:
\begin{align}
    v &= \frac{Ja_0}{2}\frac{1}{1-\beta^2}\sin[\pi(1-\beta^2)],
    \label{v_XXZ} \\
    K &= \frac{1}{2\beta^2},
    \label{K_XXZ}
\end{align}
where $\Delta = -\cos(\pi\beta^2)$.
We can also write down the Hamiltonian using the imaginary-time-dependent field, $\phi(\tau,x)$:
\begin{align}
    \mathcal H_{\rm TL}^\ast &= \frac{v}{2\pi K} \int_0^L dx \biggl( \frac{1}{v^2}(\partial_\tau\phi)^2 + (\partial_x \phi)^2 \biggr).
    \label{H_XXZ_fixedpoint_tau}
\end{align}

The asterisk on the left hand side of Eqs.~\eqref{H_XXZ_fixedpoint} and \eqref{H_XXZ_fixedpoint_tau} is put to emphasize the fact that the Hamiltonian $\mathcal H_{\rm TL}^\ast$ is valid at the fixed point in the RG flow.
The Hamiltonian \eqref{H_XXZ} deviates from the fixed-point one \eqref{H_XXZ_fixedpoint},
\begin{align}
    \mathcal H_{\rm XXZ} &= \mathcal H_{\rm TL}^\ast + \mathcal H',
    \label{H_XXZ_eff}
\end{align}
The deviation $\mathcal H'$ is irrelevant in the RG sense.
The interaction $\mathcal H'$ is marginal for $\Delta=1$ and irrelevant for $-1<\Delta<1$ and thus has no impact on the ground state although it gives logarithmic corrections to observables for $\Delta \to 1$~\cite{eggert_sus_log}.
In this section,
we neglect $\mathcal H'$ as usual for the moment and put it back when needed.

The bosonization formula \eqref{Sz2phi} transforms the polarization amplitude \eqref{def_Uq} into
\begin{align}
    U^q
    &= \exp\biggl( i\frac{2q}{L} \int_0^L dx \, x\partial_x \phi(x) \biggr)
    \notag \\
    &= \exp\biggl(i2q \phi(L) - i\frac{2q}{L}\int_0^L dx \, \phi(x) \biggr).
    \label{Uq_phi}
\end{align}
Here, we dropped the oscillating term proportional to $(-1)^j$ in Eq.~\eqref{Sz2phi} while it was taken into account in Ref.~\cite{kobayashi_polarization}.
This is because the dropped term gives rise to a subleading correction to the power law.
The ground-state average of $U^q$ has been well investigated in one-dimensional gapped quantum spin systems such as an $S=1$ Heisenberg antiferromagnetic spin chain~\cite{nakamura-todo}.
It is recognized that the polarization amplitude \eqref{def_zq} qualifies as an order parameter of valence-bond-solid phases~\cite{nakamura-todo}.
In the gapless phase of the $S=1/2$ XXZ chain, $z^{(q)}$ was calculated in Ref.~\cite{kobayashi_polarization} and turned out to be~\footnote{In Ref.~\cite{kobayashi_polarization}, the power on the right hand side was given by $2q^2K$ but it is rather by $q^2K$. This is because the power equals to the scaling dimension of the operator $e^{i2q\phi}$, which is $q^2K$.}
\begin{align}
    z^{(q)} &\propto \biggl(\frac{2\pi }{L} \biggr)^{q^2K}
    \label{scaling_zq_naive}
\end{align}
The scaling law \eqref{scaling_zq_naive} was derived using CFT by regarding $z^{(q)}$ as a multi-point correlation function of vertex operators, $e^{i\alpha\phi}$.
However, the scaling law \eqref{scaling_zq_naive} differs from numerical estimations, $z^{(2)} \propto (2\pi/L)^{4K-2}$ and $z^{(4)} \propto (2\pi/L)^{8K-4}$~\cite{kobayashi_polarization}.

Let us construct an alternative theoretical approach to the scaling law consistent with the numerical estimations.
We note that the average \eqref{def_zq} is an inner product of the ground state $\ket{\psi_0}$ and another state $U^q\ket{\psi_0}$.
The latter is an eigenstate of another Hamiltonian,
\begin{align}
    \Tilde{\mathcal H}_{\rm XXZ} := U^q\mathcal H_{\rm XXZ} U^{-q}.
    \label{tildeH_XXZ}
\end{align}
$U^q$ is a large gauge transformation for $q\in \mathbb Z$.
The transformation \eqref{tildeH_XXZ} by $U^q$ is also understood as an insertion of U(1) flux~\cite{oshikawa_lsm_flux}.

While the complete set of the eigenstates of $\tilde{\mathcal{H}}_{\rm XXZ}$ is identical to that of $\mathcal H_{\rm XXZ}$, the state $U^q\ket{\psi_0}$ is not identical to $\ket{\psi_0}$.
The ground state $\ket{\psi_0}$ equals to the vacuum state $\ket{0}_0$ of the fixed-point Hamiltonian $\mathcal H_{\rm TL}^\ast$ if we identify $\mathcal H_{\rm XXZ} = \mathcal H_{\rm TL}^\ast$.
Then, $U^q\ket{\psi_0}$ is the vacuum state $\ket{0}_q$ of the gauge-transformed Hamiltonian \eqref{tildeH_XXZ}.
The polarization amplitude $z^{(q)}$ measures the overlap of those two vacuum states:
\begin{align}
    z^{(q)} &= {}_0\braket{0|0}_q.
    \label{zq_overlap}
\end{align}
To understand this quantity, we need to relate two vacua to each other.

Let us represent $\ket{0}_q$ in terms of the original model without flux.
If we identify $\mathcal H_{\rm XXZ}$ with the fixed-point one, $\mathcal H_{\rm TL}^\ast$ [Eq.~\eqref{H_XXZ_fixedpoint}], the low-energy form of $\Tilde{\mathcal H}_{\rm XXZ}$ is derived in the following manner.
First we note that $U^q$ shifts $\partial_x \theta$ by a certain amount:
\begin{align}
    U^q \partial_x \theta (x) U^{-q} 
    &=\partial_x\theta(x) + \frac{2\pi q}{L} -2\pi q \delta(x-L).
    \label{Uq-on-theta}
\end{align}
Equation~\eqref{Uq-on-theta} follows immediately from Eq.~\eqref{Uq_phi} and the commutation relation \eqref{comm_phi}.
The integral part of the large gauge transformation $U^q$ in Eq.~\eqref{Uq_phi} increases
the winding number by $2\pi q$.
This is because we can eliminate the term $2\pi q/L$ in Eq.~\eqref{Uq-on-theta} by modifing $\theta(x)$ into $\theta'(x)$,
\begin{align}
    \theta'(x)  = \theta(x) +\frac{2\pi q}{L}x.
\end{align}
This modification keeps the periodic boundary condition intact for $q\in \mathbb Z$.
The large gauge transformation adds the winding number $2\pi q$ to $\theta(x)$:
\begin{align}
    \theta'(x+L) &= \theta'(x) + 2\pi N_\theta +2\pi q.
    \label{tbc_theta}
\end{align}
On the other hand, operation of $\exp[i2q\phi(L)]$ on $\partial_x\theta$ yields the delta function in Eq.~\eqref{Uq-on-theta}.
The role of the delta function is clarified when we move to the imaginary-time formalism.

The low-energy effective form of
the transformed Hamiltonian \eqref{tildeH_XXZ} is
\begin{align}
    \Tilde{\mathcal H}_{\rm XXZ}
    &= \frac{v}{2\pi} \int_0^L dx \biggl[ K\bigl\{\partial_x\theta' -2\pi q \delta(x-L) \bigr\}^2 + \frac{1}{K}(\partial_x \phi)^2 \biggr].
\end{align}
Regarding the canonical conjugate of $\partial_t \phi$ as $\partial_x \theta'/\pi$, we obtain the Lagrangian,
\begin{align}
    \Tilde{\mathcal L}_{\rm XXZ}
    &:= \int_0^L dx \, \partial_t \phi \frac{\partial_x\theta'}{\pi} - \Tilde{\mathcal H}_{\rm XXZ}
    \notag \\
    &= \frac{v}{2\pi K} \int_0^L dx \biggl( \frac{1}{v^2}(\partial_t\phi)^2 - (\partial_x \phi)^2 \biggr)
    \notag \\
    &\qquad +2q \partial_t \phi(t,x=L),
\end{align}
which immediately leads to
\begin{align}
    \Tilde{\mathcal H}_{\rm XXZ}
    &= \frac{v}{2\pi K} \int_0^L dx \biggl( \frac{1}{v^2}(\partial_\tau\phi)^2 + (\partial_x \phi)^2 \biggr)
    \notag \\
    &\qquad -i2 q\partial_\tau \phi(\tau,x=L).
    \label{tildeH_XXZ_tau}
\end{align}
The last term inserts vertex operators $e^{\pm i2q \phi}$ at $\tau=\pm \infty$.
In fact, the partition function $Z= \int \mathcal D \phi \exp(-\int_{-\infty}^\infty d\tau \Tilde{\mathcal H}_{\rm XXZ})$ is written as
\begin{align}
    Z &= \int \mathcal D \phi e^{i2q\phi(\infty,L)} e^{-\int_{-\infty}^\infty d\tau \mathcal{H}_{\rm TL}^\ast} e^{-i2q\phi(-\infty,L)}.
    \label{Z_tildeH_XXZ}
\end{align}

The partition function \eqref{Z_tildeH_XXZ} provides us translation rules from eigenstates in the presence of the flux to those in the absence of the flux.
Let us denote the highest-weight state of $\mathcal H_{\rm TL}^\ast$ corresponding to the operator $e^{i2n\phi}$ as 
$\ket{e^{i2n\phi}}_0$,
which is defined as~\footnote{The state $\ket{e^{i2n\phi}}_0$ is more naturally defined on the complex plane as $\ket{e^{i2n\phi}}_0=\lim_{z,\bar z \to 0}e^{i2n\phi(z,\bar z)} \ket{0}$.
The factor $(2\pi/L)^{x_{2n}} e^{-2\pi x_{2n}\beta/L}$ in Eq.~\eqref{ket_vertex} results from the conformal transformation $z=e^{2\pi(\tau+ix)/L}$ and $\bar z=e^{2\pi (\tau-ix)/L}$ from the complex plane to the cylinder.
}
\begin{align}
    \ket{e^{i2n\phi}}_0 
    &:= \lim_{T \to +\infty} \biggl(\frac{2\pi}{L}\biggr)^{x_{2n}} e^{-\frac{2\pi x_{2n}}{L}T} e^{i2n\phi(x,-T)} \ket{0}_0.
    \label{ket_vertex}
\end{align}
Here, $x_{2n}$ is the scaling dimension of vertex operators $e^{\pm i 2n\phi}$,
\begin{align}
    x_{2n} &= n^2K.
    \label{x_2n}
\end{align}
$e^{i2q\phi}$ is an operator going with a U(1) charge proportional to $q$ in the two-dimensional Coulomb gas picture~\cite{cft_yellowbook}.

The insertion of the vertex operators at $\tau=\infty$ and $\tau = -\infty$ into the partition function \eqref{Z_tildeH_XXZ} immediately means that $\ket{0}_q$ is equivalent to~\cite{kitazawa_tbc}
\begin{align}
    \ket{0}_q &= e^{i\Theta_q}\ket{e^{i2q\phi}}_0,
    \label{0q}
\end{align}
with $\Theta_q \in \mathbb R$ and 
\begin{align}
    {}_q\bra{0} &= e^{-i\Theta_q} {}_0\bra{e^{i2q\phi}}.
\end{align}
The left hand side of Eq.~\eqref{0q} is nothing but $U^q\ket{\psi_0}$.
Thus, $\Theta_q$ is proportional to $q$:
\begin{align}
    U^q\ket{\psi_0} &= e^{iq\Theta_1} \ket{e^{i2q\phi}}_0,
    \label{Uq-on-psi0}
\end{align}
The phase $\Theta_1$ is determined in accordance with symmetries that $\mathcal{H}_{\rm XXZ}$ and $\tilde{\mathcal{H}}_{\rm XXZ}$ share in common.
The simplest example is 
the one-site translation symmetry,
\begin{align}
    T_1 \bm S_j T_1^{-1} &= \bm S_{j+1},
\end{align}
or
\begin{align}
    T_1 \phi(\tau,x) T_1^{-1} &= \phi(\tau,x) + \frac{\pi}{2}.
    \label{def_T1}
\end{align}
Both $\mathcal H_{\rm XXZ}$ and $\tilde{\mathcal H}_{\rm XXZ}$ have the one-site translation symmetry.
However, $U^q$ and $T_1$ do not commute with each other in general~\cite{oshikawa_lsm_flux}.
In fact, 
\begin{align}
    T_1 U^q T_1^{-1} &= e^{i\pi q} U^q,
    \label{T1-Uq}
\end{align} 
holds true for the $S=1/2$ spin chains.

Thanks to Eq.~\eqref{T1-Uq},
$\mathcal{H}_{\rm XXZ}$ and $\tilde{\mathcal{H}}_{\rm XXZ}$ are both symmetric under the one-site translation.
The ground state is an eigenstate of $T_1$ with an eigenvalue $e^{iP_0}$,
\begin{align}
    T_1\ket{\psi_0} = e^{iP_0} \ket{\psi_0},
\end{align}
which can be rephrased as
$T_1\ket{0}_0 = e^{iP_0}\ket{0}_0$.
Operation of $T_1$ on Eq.~\eqref{Uq-on-psi0} results in
$\Theta_1=0 \mod 2\pi$ .
We thus end up with
\begin{align}
    z^{(q)}
    &= {}_0\braket{0|e^{-i2q\phi}}_0.
    \label{zq_overlap_0}
\end{align}
As far as $\mathcal H_{\rm XXZ}$ is identified with $\mathcal H_{\rm TL}^\ast$, the right hand side is zero for many reasons such as the charge neutrality~\cite{cft_yellowbook}.

We found that we need to take the perturbation $\mathcal H'$ in the Hamiltonian \eqref{H_XXZ_eff} into account.
Unless the parameter is fine tuned, the Hamiltonian of the effective field theory usually deviates from that at the fixed-point.
The $S=1/2$ XXZ spin chain is a typical example.
The most relevant interaction in $\mathcal H'$ [Eq.~\eqref{H_XXZ_eff}] for $\Delta \simeq 1$ is the umklapp term~\footnote{Strictly speaking, $\cos(4\phi)$ with the scaling dimension $4K$ is llessess relevant than the marginal interaction $(\partial_x\phi)^2-(\partial_x\theta)^2$. However, this interaction is absorbed into the fixed-point Hamiltonian as a renormalization of the velocity $v$ and thus discarded here.},
\begin{align}
    \mathcal H' &= \lambda \int_0^L \frac{dx}{2\pi} \cos(4\phi).
    \label{H'_4phi}
\end{align}
The importance of the umklapp interaction \eqref{H'_4phi} was already pointed out in Ref.~\cite{kobayashi_polarization}.
However, it remains obscure how we should take properly the umklapp term into account.

The vacua $\ket{0}_0$ and $\ket{0}_q$ are perturbed by $\mathcal H'$ and
\begin{align}
     \tilde{\mathcal H}':=U^q\mathcal H' U^{-q},
     \label{tH'_def}
\end{align}
respectively.
The transformed perturbation \eqref{tH'_def} plays the central role in discussions in the subsequent sections.
The perturbative expansions of $\ket{0}_0$ and $\ket{0}_q$ are given in the following well-knonwn form:
\begin{widetext}
\begin{align}
    \ket{0}_0 &\longrightarrow \ket{0}_0 - \sum_{n(\not=0)} \frac{{}_0\braket{e^{i2n\phi}|\mathcal H'|0}_0}{E_{2n}-E_0}\ket{e^{i2n\phi}}_0 
    \notag \\
    &\qquad + \sum_{n(\not=0)} \sum_{m(\not=0)} \frac{{}_0\braket{e^{i2m\phi}|\mathcal H'|e^{i2n\phi}}_0{}_0\braket{e^{i2n\phi}|\mathcal H'|0}_0}{(E_{2n}-E_0)(E_{2m}-E_0)} \ket{e^{i2m\phi}}_0 
     - \sum_{n(\not=0)} \frac{{}_0\braket{e^{i2n\phi}|\mathcal H'|0}_0 {}_0 \braket{0|\mathcal H'|0}_0}{ (E_{2n}-E_0)^2}
    + \cdots,
    \label{0_perturbation} \\
    \ket{0}_q
    &\longrightarrow \ket{0}_q - \sum_{n(\not=0)} \frac{{}_q\braket{e^{i2n\phi}|\tilde{\mathcal H}'|0}_q}{E_{2n}-E_0}\ket{e^{i2n\phi}}_q
    \notag \\
    &\qquad + \sum_{n(\not=0)} \sum_{m(\not=0)} \frac{{}_q\braket{e^{i2m\phi}|\tilde{\mathcal H}' |e^{i2n\phi}}_q{}_q\braket{e^{i2n\phi}|\tilde{\mathcal H}'|0}_q}{(E_{2n}-E_0)(E_{2m}-E_0)} \ket{e^{i2m\phi}}_q 
     - \sum_{n(\not=0)} \frac{{}_q\braket{e^{i2n\phi}|\tilde{\mathcal H}'|0}_q {}_q \braket{0|\tilde{\mathcal H}'|0}_q}{ (E_{2n}-E_0)^2}
    + \cdots, \notag \\
    &=\ket{e^{i2q\phi}}_0 - \sum_{n(\not= 0)} \frac{{}_0\braket{e^{i2(n+q)\phi}|\tilde{\mathcal H}'|e^{i2q\phi}}_0}{E_{2n}-E_{0}} \ket{e^{i2(n+q)\phi}}_0
        \notag \\
    &\qquad + \sum_{n,m(\not=0)} \frac{{}_0\braket{e^{i2(m+q)\phi}|\tilde{\mathcal H}'|e^{i2(n+q)\phi}}_0{}_0\braket{e^{i2(n+q)\phi}|\tilde{\mathcal H}'|e^{i2q\phi}}_0}{(E_{2n}-E_{0})(E_{2m}-E_{0})} \ket{e^{i2(m+q)\phi}}_0 
    \notag \\
    &\qquad
   - \sum_{n(\not=0)} \frac{{}_0\braket{e^{i2(n+q)\phi}|\tilde{\mathcal H}'|e^{i2q\phi}}_0 {}_0 \braket{e^{i2q\phi}|\tilde{\mathcal H}'|e^{i2q\phi}}_0}{ (E_{2n}-E_0)^2} + \cdots.
    \label{2n_perturbation}
\end{align}
Here, $E_{0}$ and $E_{2n}$ are the unperturbed eigenenergies of $\ket{0}_0$ and $\ket{e^{\pm i2n\phi}}_0$  of $\mathcal H_{\rm TL}^\ast$, respectively.
We emphasize that $\ket{0}_q$ and $\ket{e^{\pm i2n\phi}}_q$ also possess the unperturbed eigenenergies $E_0$ and $E_{2n}$ of the transformed Hamiltonian $U^q\mathcal H_{\rm TL}^\ast U^{-q}$, respectively.
The excitation gap $E_{2n}-E_0$ is proportional to $2\pi/L$:
\begin{align}
    E_{2n}-E_0 = \frac{2\pi}{L} x_{2n}.
\end{align}
We discarded in Eqs.~\eqref{0_perturbation} and \eqref{2n_perturbation} eigenstates which contain spatial or temporal derivatives.
This is because the perturbation \eqref{H'_4phi} does not have any derivative in the integrand.
The terms involving such derivatives give subleading contributions to the scaling law which is out of our focus in this paper.

The first-order correction to the polarization amplitude, which we denote $\delta_1 z^{(q)}$ , is thus given by
\begin{align}
    \delta_1 z^{(q)}
    &= -\biggl[\sum_{n(\not=0)} \frac{{}_0\braket{0|\mathcal H'|e^{i2n\phi}}_0}{E_{2n}-E_0} {}_0 \braket{e^{i2n\phi}|e^{i2q\phi}}_0
    + \sum_{n(\not=0)} \frac{{}_0 \braket{e^{i2(n+q)\phi}|\tilde{\mathcal H}'|e^{i2q\phi}}_0}{E_{2n
    }-E_0}{}_0\braket{0|e^{i2(n+q)\phi}}_0
    \biggr]
    \notag \\
    &\simeq -2 \frac{{}_0 \braket{0|\mathcal H'|e^{i2q\phi}}_0}{E_{2q}-E_0} .
\end{align}
In the last line, we approximated $\tilde{\mathcal H}'\simeq \mathcal H'$.
This relation is exact when $L\to + \infty$.
The difference of $\tilde{\mathcal H}'$ from $\mathcal H'$ is however nonzero at the level of the lattice model and will be discussed in the next section.

Since the matrix element ${}_0\braket{0|\mathcal H'|e^{i2q\phi}}_0$ is given by~\cite{cardy}
\begin{align}
    {}_0\braket{0|\mathcal H'|e^{i2q\phi}}_0 = \frac{\lambda L}{4\pi} \biggl(\frac{2\pi}L\biggr)^{x_4} (\delta_{q,2}+\delta_{q,-2})
\end{align}
and the precise form of the energy denominator is available with the aid of the conformal field theory, we obtain
\begin{align}
    \delta_1 z^{(q)}
    &= -\frac{\lambda}{x_4} \biggl(\frac{2\pi}{L}\biggr)^{x_4-2}(\delta_{q,2}+\delta_{q,-2}).
\end{align}
It correctly reproduces the numerical result of the scaling law~\cite{kobayashi_polarization}.

The second-order correction $\delta_2z^{(q)}$ to $z^{(q)}$ is also easily derived.
Keeping nonzero terms only, we obtain the second-order correction,
\begin{align}
    \delta_2 z^{(q)}
    &= \sum_{n(\not=0)} \frac{{}_0\braket{0|\mathcal H'|e^{i2n\phi}}_{00}\braket{e^{i2n\phi}|\mathcal H'|e^{i2q\phi}}_0}{(E_{2n}-E_0)(E_{2q}-E_0)}
    +  \sum_{n(\not=0)} \frac{{}_0\braket{0|U^q\mathcal H'U^{-q}|e^{i2(n+q)\phi}}_{00}\braket{e^{i2(n+q)\phi}|U^q\mathcal H'U^{-q}|e^{i2q\phi}}_0}{(E_{2n}-E_0)(E_{2q}-E_0)}
    \notag \\
    &\qquad 
    +\sum_{n(\not=0)} \frac{{}_0\braket{0|\mathcal H'|e^{i2n\phi}}_{00}\braket{e^{i2n\phi}|U^q\mathcal H'U^{-q}|e^{i2q\phi}}_0}{(E_{2n}-E_0)(E_{2(n-q)}-E_0)}.
\end{align}
The matrix elements in the numerators give a restriction on $q$ due to the charge neutrality, for instance,
\begin{align}
    {}_0\braket{0|\mathcal H'|e^{i2n\phi}}_0 {}_0 \braket{e^{i2n\phi}|\mathcal H'|e^{i2q\phi}}_0
    &= \frac{\lambda^2}4 \delta_{n,2}\delta_{q,4} \int_0^L \frac{dx dy}{(2\pi)^2} {}_0\braket{0|e^{-i4\phi(x)}|e^{i4\phi}}_0 {}_0\braket{e^{i4\phi}| e^{-i4\phi(y)}|e^{i2q\phi}}_0
    \notag \\
    &\qquad +\frac{\lambda^2}4 \delta_{n,-2}\delta_{q,-4} \int_0^L \frac{dxdy}{(2\pi)^2} {}_0\braket{0|e^{i4\phi(x)}|e^{-i4\phi}}_0 {}_0 \braket{e^{-i4\phi}|e^{i4\phi(y)}|e^{i2q\phi}}_0
    \notag \\
    &= \frac{\lambda^2}4 (\delta_{n,2}\delta_{q,4}+\delta_{n,-2}\delta_{q,-4}) \biggl(\frac{2\pi}{L}\biggr)^{2x_4-2}.
\end{align}
\end{widetext}
It results in the following second-order correction:
\begin{align}
    \delta_2z^{(q)} 
    &= \frac{3\lambda^2}{4x_4x_8} (\delta_{q,4} + \delta_{q,-4}) \biggl(\frac{2\pi}{L}\biggr)^{2x_4-4} .
\end{align}
Therefore, $z^{(4)}$ follows the scaling law $z^{(4)} \propto (2\pi/L)^{8K-4}$, which reproduces the numerical result~\cite{kobayashi_polarization}.

\section{Fine-tuned $J_1$-$J_2$ XXZ spin chain} \label{sec:J1J2}

In the last section, we clarified the relevance of the irrelevant interaction \eqref{H'_4phi} in the  polarization amplitude $z^{(q)}$.
The importance of the irrelevant interaction was deduced in Ref.~\cite{kobayashi_polarization} from comparison of the scaling laws in the $S=1/2$ XXZ spin chain and in the $S=1/2$ $J_1$-$J_2$ XXZ spin chain.
The latter model is described by the following Hamiltonian,
\begin{align}
    \mathcal H_{J_1-J_2} &= \sum_{n=1,2} \sum_{j=1}^L J_n (S_j^x S_{j+n}^x + S_j^y S_{j+n}^y + \Delta S_j^z S_{j+n}^z).
    \label{H_J1J2}
\end{align}
Fine tuning of the ratio $J_2/J_1$ eliminates the coupling constant of $\cos(4\phi)$ in the $L \to + \infty$ limit, 
where the system is on the quantum critical point described by the TL liquid.
The quantum critical point corresponds to the  Berezinskii-Kosterlitz-Thouless transition point for $\Delta =1$~\cite{Berezinskii, Kosterlitz_1973, Kosterlitz_1974}.
It seems that the polarization amplitude $z^{(2)}$ and $z^{(4)}$ are to be exactly zero without the interaction \eqref{H'_4phi}.
Nevertheless, scaling laws $z^{(2)} \propto (2\pi/L)^{4K}$ and $z^{(4)}\propto (2\pi/L)^{8K}$ were observed~\cite{kobayashi_polarization}.
The powers are shifted from those in the XXZ spin chain by universal amounts.

Precisely speaking, another irrelevant interaction $\cos(8\phi)$ is present even when $\lambda$ is fine-tuned to be zero.
This highly irrelevant interaction in the RG sense can make $z^{(4)}$ nonzero.
The $\cos(8\phi)$ interaction adds to $z^{(4)}$ a term proportional $(2\pi/L)^{x_8-2}$.
However, it disagrees with the numerical result~\cite{kobayashi_polarization}.

The numerically observed scaling law implies that the $\cos(4\phi)$ interaction somehow survives in the finite $L$ case because of the estimated power $\beta_2=x_4$.
To confirm the implication, we need to recall the fact that the state $U^q\ket{\psi_0}=\ket{0}_q$ is the ground state of $\Tilde{\mathcal H}_{J_1-J_2}$ but not of $\mathcal{H}_{J_1-J_2}$ at finite $L$.
The former is defined as
\begin{align}
    \Tilde{\mathcal H}_{J_1-J_2}
    &= U^q \mathcal H_{J_1-J_2} U^{-q}.
\end{align}
This transformation was discussed in Ref.~\cite{hirano_bp_J1J2} for $J_1$-$J_2$ spin-$S$ chains to evaluate the Berry phase in their ground-state phase diagrams.
The Berry phase characterizes the valence-bond-solid phase in one dimension as well as the polarization amplitude does.
This similarity is easily understood by paying attention to a fact that both discussions on the Berry phase and on the polarization amplitude rely on the transformation $U^q$.

It follows from $U^qS_j^\pm U^{-q} = e^{\pm i2\pi q j /L}S_j^\pm$ and $U^q S_j^z U^{-q} = S_j^z$ that
\begin{widetext}
\begin{align}
    \Tilde{\mathcal H}_{J_1-J_2}
    &= \sum_{n=1,2}\sum_{j=1}^L J_n \biggl( \frac 12 (e^{-\frac{i2\pi nq }L} S_j^+S_{j+n}^- + e^{i\frac{2\pi nq }L} S_j^- S_{j+n}^+) 
    + \Delta S_j^z S_{j+n}^z \biggr)
    \notag \\
    &= \mathcal H_{J_1-J_2} -\sum_{n=1,2} J_n \sin\biggl(\frac{2\pi nq }{L}\biggr) \sum_{j=1}^L (S_j^x S_{j+n}^y - S_j^y S_{j+n}^x)
   - \sum_{n=1,2} J_n \biggl[1-\cos\biggl(\frac{2\pi nq }{L}\biggr) \biggr] \sum_{j=1}^L (S_j^x S_{j+n}^x + S_j^y S_{j+n}^y).
    \label{tildeH_J1J2}
\end{align}
\end{widetext}
For $L \gg a_0$, the second term of Eq.~\eqref{tildeH_J1J2} yields the interaction proportional to $(\pi q/L) \partial_x \theta$ which is also present in the low-energy effective form of $\Tilde{\mathcal H}_{\rm XXZ}$.
The last term of Eq.~\eqref{tildeH_J1J2} was not taken into account in Eq.~\eqref{tildeH_XXZ}, which leads to the following interaction:
\begin{align}
    \tilde{\mathcal H}'
    &= \lambda' \biggl(\frac{2\pi q}{L}\biggr)^2 \int_0^L \frac{dx}{2\pi} \cos(4\phi).
    \label{H'_J1J2}
\end{align}
This observation of the finite-$L$ system makes it possible to treat the models in the last and present sections on equal footing.
At finite $L$, the irrelevant interaction \eqref{H'_4phi} is generically given by
\begin{align}
    \tilde{\mathcal H}' = \lambda_L \int_0^L \frac{dx}{2\pi} \cos(4\phi),
    \label{H'_4phi_modified}
\end{align}
with the coupling constant $\lambda_L$:
\begin{align}
    \lambda_L = \lambda + \lambda'\biggl(\frac{2\pi q}{L}\biggr)^2.
    \label{lambda_L}
\end{align}
We approximated $\lambda_L\simeq \lambda$
in the XXZ chain \eqref{H_XXZ} because $\lambda_L-\lambda=\lambda'(2\pi q/L)^2$ is negligible compared to $\lambda$ for large $L$.
However, in the $J_1$-$J_2$ XXZ chain, the fine tuning makes $\lambda$ be zero and thus $\lambda'(2\pi q/L)^2$ be the leading contribution to $\lambda_L$. 

The interaction \eqref{H'_J1J2} is irrelevant in the RG sense and negligible in the $L \to + \infty$ limit.
Even when we fine-tuned the parameters to realize $\lambda_\infty=0$,  the coupling constant $\lambda_L$ is nonzero for finite $L$ and the interaction \eqref{H'_4phi_modified} perturbs the state $\ket{0}_q$.
The state $\ket{0}_q$ gets perturbed by $\tilde{\mathcal H}'$ [Eq.~\eqref{2n_perturbation}].
The matrix element ${}_0 \braket{e^{i2(n+q)\phi}|\tilde{\mathcal H}'|e^{i2q\phi}}_0$ in the first-order perturbation term is nonzero for $n=\pm 2$:
\begin{align}
    {}_0\braket{e^{i2(n+q)\phi} |\tilde{\mathcal H}'|e^{i2q\phi}}_0
    &= \frac{q^2\lambda'}{2} (\delta_{n,2}+\delta_{n,-2}) \biggl(\frac{2\pi}{L}\biggr)^{x_4+1}.
\end{align}
The first-order correction  immediately turns out to be
\begin{align}
    \delta_1 z^{(q)}
    &= -2 (\delta_{q,2}+\delta_{q,-2}) \frac{{}_0\braket{0|\tilde{\mathcal H}'|e^{i2q\phi}}_0}{E_{2q}-E_0}
    \notag \\
    &=-\frac{q^2\lambda'}{x_4}(\delta_{q,2}+\delta_{q,-2}) \biggl(\frac{2\pi}{L}\biggr)^{x_4}.
    \label{zq_J1J2}
\end{align}
This result explains the numerically obtained scaling law $z^{(2)} \propto (2\pi/L)^{x_4}$~\cite{kobayashi_polarization}.
The extra factor $(2\pi/L)^2$ in Eq.~\eqref{H'_J1J2} shifts the power.
Similarly, we obtain $z^{(4)} \propto (2\pi/L)^{2x_4}$ from the second-order perturbation, which is again consistent with the numerical estimation.

\section{Haldane-Shastry model}\label{sec:HS}

The fine tuning was required in the spin-$1/2$ $J_1$-$J_2$ XXZ spin chain to eliminate the $\cos(4\phi)$ interaction in the $L\to + \infty$ limit.
In contrast, the Haldane-Shastry model is known to give the TL liquid ground state described by the fixed-point Hamiltonian with $K=1/2$ and $\mathcal H'=0$ requiring no fine tuning~\cite{haldane_HS, shastry_HS}.
The Haldane-Shastry model has the following long-range interaction:
\begin{align}
    \mathcal{H}_{\rm HS} &= \frac{J}{4}\biggl(\frac{2\pi}{L}\biggr)^2 \sum_{r<r'} \frac{\bm S_r \cdot \bm S_{r'}}{\sin^2[\pi (r'-r)/L]}.
\end{align}
The ground state of the Haldane-Shastry model is exactly $\ket{0}_0$ even for finite $L$.
To evaluate finite-size effects on $\ket{0}_q$, we transform the Haldane-Shastry model by $U^q$.
\begin{widetext}
\begin{align}
    \Tilde{\mathcal H}_{\rm HS}
    &= U^q \mathcal H_{\rm HS} U^{-q}
    \notag \\
    &= \mathcal H_{\rm HS}
    - \frac{J}{4}\biggl(\frac{2\pi}{L}\biggr)^2  \sum_{r<r'} \frac{\sin[2\pi q(r'-r)/L]}{\sin^2[\pi (r'-r)/L]} (S_r^xS_{r'}^y - S_r^yS_{r'}^x)
    \notag \\
    &\quad - \frac{J}{4}\biggl(\frac{2\pi}{L}\biggr)^2 \sum_{r<r'} \frac{1-\cos[2\pi q(r'-r)/L]}{\sin^2[\pi(r'-r)/L]} (S_r^x S_{r'}^x + S_r^y S_{r'}^y).
    \label{tildeH_HS}
\end{align}
\end{widetext}
In analogy with the $J_1$-$J_2$ XXZ spin chain, we may expect that the last line of Eq.~\eqref{tildeH_HS}, which we denote as $\tilde{\mathcal H}'_{\rm HS}$, makes ${}_0\braket{0|\tilde{\mathcal H}'_{\rm HS}|e^{-i4\phi}}_0$ nonzero.
However, $\tilde{\mathcal H}'_{\rm HS}$ cannot be simply reduced to Eq.~\eqref{H'_4phi_modified} because of its long-range nature.
To understand the perturbation,
\begin{align}
    \tilde{\mathcal H}'_{\rm HS}
    &=\sum_{r,r'} J'_{r,r'}(S_r^x S_{r'}^x + S_r^y S_{r'}^y),
    \label{H'_HS}
\end{align}
with
\begin{align}
    J'_{r,r'}
    &=- \frac{J}{4}\biggl(\frac{2\pi}{L}\biggr)^2  \frac{1-\cos[2\pi q(r'-r)/L]}{\sin^2[\pi(r'-r)/L]},
    \label{J'_HS}
\end{align}
we first closely look into the coupling constant \eqref{J'_HS}.
Using the Chebyshev polynomial of the second kind $u_n(x)$, we can express it as
\begin{align}
    J'_{r,r'} &= - \frac{J}{4}\biggl(\frac{2\pi}{L}\biggr)^2   u_{q-1}\Bigl(\cos\tfrac{\pi(r'-r)}L \Bigr)^2.
    \label{J'_HS_chebyshev}
\end{align}
The polynomial $u_n(x)$ is defined for $n\in \mathbb Z$ as
\begin{align}
    u_n(\cos\theta) := \frac{\sin[(n+1)\theta]}{\sin\theta}.
\end{align}
It is a family of the Chebyshev polynomial $t_n(x)$ (of the first kind),
\begin{align}
    t_n(\cos\theta) := \cos(n\theta).
\end{align}
There are identities relating two Chebyshev polynomials.
For example, the following relation holds true for even $n$:
\begin{align}
    u_n(x) &= -1+2\sum_{m=0}^{n/2} t_{2m}(x).
\end{align}
Combining it with another identity $[u_n(x)]^2 = \sum_{m=0}^n u_{2m}(x)$, we obtain
\begin{align}
    [u_{q-1} (x)]^2
    &= -q + 2\sum_{m=0}^{q-1}\sum_{l=0}^{m}t_{2l}(x).
\end{align}
In other words,
\begin{align}
    J'_{r,r'}
    &=  \frac{J}{4}\biggl(\frac{2\pi}{L}\biggr)^2 \biggl(q - \sum_{m=0}^{q-1}\sum_{l=0}^{m} (e^{i\frac{2\pi l}L(r'-r)}+e^{-i\frac{2\pi l}L(r'-r)}) \biggr). 
    \label{J'_HS_sum}
\end{align}
Equation~\eqref{J'_HS_sum} is nothing but the Fourier transformation of $J'_{r,r'}$.
The coefficient $J'_{r,r'}$ turned out to contain the uniform part coupled to $q J(2\pi/L)^2/4$ and the other oscillating part.

For the moment, we focus on the following perturbation instead of dealing with $\tilde{\mathcal H}'_{\rm HS}$ directly,
\begin{align}
    \mathcal H'_{k,k'} &= \lambda_{k,k
    '} \biggl(\frac{2\pi}{L} \biggr)^2 \int_0^L \frac{drdr'}{(2\pi)^2} e^{ikr} e^{ik'r'}
    \notag \\
    &\qquad \times [ J_R^x(r) J_L^x(r') + J_R^y(r) J_L^y(r')],
    \label{H'_kk'}
\end{align}
where $k,k'\in [0, 2\pi/L)$ and $J_R^a(r)$ and $J_L^a(r)$ ($a=x,y$) are the right-moving and the left-moving parts of the uniform component of $S_r^a$.
For $S_r^\pm =S_r^x \pm i S_r^y$, they are
\begin{align}
    J_R^\pm (r) &= e^{\pm i[2\phi(r)+\theta'(r)]}, \\
    J_L^\pm (r) &= e^{\mp i[2\phi(r)-\theta'(r)]}.
\end{align}
$J_R^a$ and $J_L^a$ have the conformal weights $(x_4/2,0)$ and $(0,x_4/2)$, respectively.
$\tilde{\mathcal H}'_{\rm HS}$ is written as a superposition of Eq.~\eqref{H'_kk'} with several $k$ and $k'$.

At the fixed-point, the right-moving and the left-moving parts are decoupled.
Thus, the highest-weight state $\ket{e^{i4\phi}}_0$ is a product of two states in right-moving and left-moving parts,
\begin{align}
    \ket{e^{i4\phi}}_0 &= \ket{J_R^+}_0 \ket{J_L^-}_0.
\end{align}
The matrix element ${}_0 \braket{0|\tilde{\mathcal H}'_{\rm HS}|e^{i4\phi}}_0$ is thus analytically obtained,
\begin{widetext}
\begin{align}
    {}_0 \braket{0|\mathcal H'_{k,k
    '}|e^{i4\phi}}_0
    &= q\lambda''\biggl(\frac{2\pi}{L}\biggr)^2\sum_{a=x,y} \int_0^L \frac{dr}{2\pi}  {}_0 e^{ikr} \braket{0|J_R^a(r)|J_R^+}_0
    \int_0^L \frac{dr'}{2\pi} e^{ik'r'}\braket{0|J_L^a(r')|J_L^-}_0
    \notag \\
    &= 2q\lambda'' \biggl(\frac{2\pi}{L} \biggr)^{x_4}\delta_{k,0}\delta_{k',0},
\end{align}
\end{widetext}
thanks to the relation ${}_0\braket{0|J_{R/L}^\pm(r)|J_{R/L}^\mp}_0 = 2(2\pi/L)^{x_4/2}$ which holds true independent of $r$~\cite{cardy}.
The scaling law in the Haldane-Shastry model is thus determined only by the uniform part of the long-range irrelevant interaction $\tilde{\mathcal H}'_{\rm HS}$:
\begin{align}
    \tilde{\mathcal H}'_{\rm HS} &= q\lambda'' \biggl(\frac{2\pi}{L}\biggr)^2 \int_0^L \frac{drdr'}{(2\pi)^2}  [ J_R^x(r) J_L^x(r') + J_R^y(r) J_L^y(r')]
    \notag \\
    &\qquad + \sum_{k(\not=0)}\sum_{k'(\not=0)} \mathcal H_{k,k'},
    \label{H'_HS_eff}
\end{align}
with $\lambda'' = O(J)$.
The scaling law is given by
\begin{align}
    z^{(2)} =- \frac{q\lambda''}{2x_4} \biggl(\frac{2\pi}{L}\biggr)^{x_4-1}.
    \label{z2_HS}
\end{align}
The power $x_4-1=1$ is identical to that coming out of the Gutzwiler-Jastrow wave function~\cite{kobayashi_polarization}.
It is obvious that the second-order perturbation results in $z^{(4)} \propto (2\pi/L)^{2(x_4-1)} = (2\pi/L)^2$.

The reduction of the power $\beta_2=x_4-1$ by $1$ from $\beta_2=x_4$ for the $J_1$-$J_2$ XXZ chain is clarified by comparing Eq.~\eqref{H'_HS_eff} with the corresponding interaction \eqref{H'_J1J2} in the $J_1$-$J_2$ XXZ chain.
The latter is also written as
\begin{align}
    \tilde{\mathcal H}'= q^2\lambda' \biggl(\frac{2\pi}{L}\biggr)^2\int_0^L \frac{dr}{2\pi} [J_R^x(r)J_L^x(r) + J_R^y(r) J_L^y(r)].
    \label{H'_J1J2_JJ}
\end{align}
Since the matrix element ${}_0\braket{0|J_R^a(r)J_L^a(r')|e^{i4\phi}}_0 = {}_0\braket{0|J_R^a(r)|J_R^+}_{00}\braket{0|J_L^a(r')|J_L^-}_0$ is independent of $r$ and $r'$, the matrix element ${}_0\braket{0|\tilde{\mathcal H}'_{\rm HS}|e^{i4\phi}}_0$ acquired the factor $L^2$ coming out of the spatial integrals.
On the other hand, ${}_0\braket{0|\mathcal H'|e^{i4\phi}}_0$ for Eq.~\eqref{H'_J1J2_JJ} acquired the factor $L$ from the spatial integral.
The difference purely results from the  range of interaction.

\section{$S=1/2$ spin ladder}\label{sec:ladder}

In the previous Secs.~\ref{sec:XXZ}, \ref{sec:J1J2}, and \ref{sec:HS}, we dealt with the ground state of three models which is nontrivial in the sense of the LSM theorem~\cite{lsm}.
Namely, we considered cases where the fixed-point theory is exposed only to irrelevant interactions.
In this section, we discuss effects of a relevant interaction to the fixed-point theory of the TL liquid.

We consider an $S=1/2$ spin ladder
\begin{align}
    \mathcal H_{\rm ladder}
    &= J \sum_{j=1}^L \sum_{n=1,2} \bm S_{j,n} \cdot \bm S_{j+1,n}
    \notag \\
    &\qquad +\sum_j [ J_\perp \bm S_{j,1} \cdot \bm S_{j,2}
    \notag \\
    &\qquad +J_\times (\bm S_{j,1} \cdot \bm S_{j+1,2} + \bm S_{j,2} \cdot \bm S_{j+1,1})
    ],
    \label{H_ladder}
\end{align}
with $J$ and $0<\max\{|J_\perp|, |J_\times|\} \ll J$.
When $J_\perp$ and $J_\times$ satisfy
\begin{align}
    J_\perp = J_\times,
    \label{ladder_qcp}
\end{align}
the excitation gap is closed~\cite{kim_ladder}.
The polarization amplitude was also discussed in the spin ladder \eqref{H_ladder} in Refs.~\cite{nakamura_ptp, nakamura_twist_physica}.

The low-energy effective field theory exactly at the quantum critical point \eqref{ladder_qcp} is the TL liquid \eqref{H_XXZ_fixedpoint}.
We bosonize the spin ladder based on
\begin{align}
    S_{j,n}^z &= \frac{a_0}{\pi}\partial_x \phi_n + (-1)^{j+n} a_1 \cos(2\phi_n), \\
    S_{j,n}^\pm &= e^{\pm i\theta_n} \bigl[ (-1)^{j+n}b_0 + b_1 \cos(2\phi_n) \bigr].
\end{align}
We recombine $\phi_n$ and $\theta_n$ ($n=1,2$) and deal with symmetric and antisymmetric modes:
\begin{align}
    \phi_\pm &:= \frac{\phi_1 \pm \phi_2}{\sqrt 2}, \qquad 
    \theta_\pm := \frac{\theta_1 \pm \theta_2}{\sqrt 2}.
\end{align}
It is well known that
only the $\phi_+$ mode participates in
the quantum phase transition at the parameter \eqref{ladder_qcp} and the $\phi_-$ degree of freedom contributing only to higher-energy gapped modes is negligible.
The low-energy Hamiltonian of the ladder \eqref{H_ladder} near the transition point is sine-Gordon-type one.
\begin{align}
    \mathcal{H}_{\rm ladder}
    &= \frac{v}{2\pi}\int_0^L dx \biggl( K(\partial_x\theta_+)^2 + \frac 1K (\partial_x \phi_+)^2 \biggr)
    \notag \\
    &\qquad + g \int_0^L \frac{dx}{2\pi} \cos (2\sqrt 2\phi_+).
    \label{H_ladder_eff}
\end{align}
where the coupling $g\propto (J_\perp - J_\times)$ vanishes at the transition point.

Let us perform $U^q$ on $\mathcal H_{\rm ladder}$.
We define $U^q$ in the spin ladder as
\begin{align}
    U^q 
    &:= \exp\biggl(i\frac{2\pi q}{L} \sum_{j=1}^L j(S_{j,1}^z+S_{j,2}^z) \biggr)
    \notag \\
    &= \exp\biggl(i\frac{2\sqrt 2 q}{L} \phi_+(L) - i\frac{2\sqrt 2 q}{L} \int_0^L dx \phi_+(x) \biggr).
    \label{Uq_ladder}
\end{align}
Note that the operator $U^q$ does not include the antisymmetric mode by definition.
At the level of the effective field theory, $U^q\mathcal H_{\rm ladder} U^{-q} =: \tilde{\mathcal H}_{\rm ladder}$ acquires the shift of $\partial_x \theta_+$,
\begin{align}
    \tilde{\mathcal H}_{\rm ladder}
    &= \frac{v}{2\pi}\int_0^L dx \biggl( K\biggl(\partial_x\theta_+ + \frac{2\sqrt 2\pi q}{L} -2\sqrt 2\pi q \delta(x-L) \biggr)^2
    \notag \\
    &\qquad + \frac{1}{K}(\partial_x \phi_+)^2 \biggr),
\end{align}
which is equivalent to
\begin{align}
    \tilde{\mathcal{H}}_{\rm ladder}
    &= \frac{v}{2\pi K}\int_0^L dx \biggl( \frac 1{v^2}(\partial_\tau \phi)^2 + (\partial_x \phi)^2 \biggr) 
    \notag \\
    &\qquad -i2\sqrt 2q \partial_\tau \phi(\tau,x=L).
\end{align}
The last term means that $\ket{0}_q$ of $\tilde{\mathcal H}_{\rm ladder}$ equals to $\ket{e^{i2\sqrt 2q\phi_+}}_0$,
\begin{align}
    \ket{0}_q = \ket{e^{i2\sqrt 2q\phi_+}}_0.
\end{align}
The phase factor is determined in the same manner with the spin chains.

Up to the first-order perturbation expansion, the polarization amplitude is given by
\begin{align}
    z^{(q)}
    &\simeq
    - \frac{{}_0\braket{0 |\tilde{\mathcal H}'|e^{i2\sqrt 2q\phi}}_0}{E_{2\sqrt 2q}-E_0}.
    \label{zq_ladder_formula}
\end{align}
$\tilde{\mathcal H}'$ in Eq.~\eqref{zq_ladder_formula} is the perturbation to the fixed-point Hamiltonian in the gauge-transformed system.
At the level of the lattice model, the Hamiltonian of the spin ladder is transformed into
\begin{widetext}
\begin{align}
    \tilde{\mathcal{H}}_{\rm ladder}
    &= U^q \mathcal{H}_{\rm ladder} U^{-q}
    \notag \\
    &= \mathcal{H}_{\rm ladder}
    +J\sin\biggl(\frac{2\pi q}{L}\biggr) \sum_j \sum_{n=1,2}
    (S_{j,n}^xS_{j+1,n}^y - S_{j,n}^y S_{j+1,n}^x)
    \notag \\
    &\qquad +J_\perp \sin\biggl(\frac{2\pi q}{L}\biggr) \sum_j (S_{j,1}^xS_{j+1,2}^y - S_{j,1}^y S_{j+1,2}^x 
    +S_{j,2}^xS_{j+1,1}^y - S_{j,2}^y S_{j+1,1}^x)
    \notag \\
    &\qquad
    - 2J \sin^2\biggl(\frac{\pi q}{L}\biggr)
    \sum_j \sum_{n=1,2} (S_{j,n}^xS_{j+1,n}^x +  S_{j,n}^yS_{j+1,n}^y)
    \notag \\
    &\qquad 
    -2J_\perp\sin^2\biggl(\frac{\pi q}{L}\biggr)  \sum_j  (S_{j,1}^xS_{j+1,2}^x +  S_{j,1}^yS_{j+1,2}^y
    +S_{j,2}^xS_{j+1,1}^x +  S_{j,2}^yS_{j+1,1}^y)
\end{align}
\end{widetext}
The terms proportional to $\sin(2\pi q/L)$ are regarded as interactions $(2\pi q/L)\partial_x \theta_+$ and $(2\pi q/L)\partial_x\theta_-$ with certain coupling constants, which are insignificant in evaluating the polarization amplitude.
The terms proportional to $\sin^2(\pi q/L)$ are essential.
They are turned into a relevant interaction $(\pi q/L)^2 \cos(2\sqrt 2\phi_+)$ at low energies.
Let us write the relevant interaction $\tilde{\mathcal H}'$ in $\tilde{\mathcal{H}}_{\rm ladder}$ as
\begin{align}
    \tilde{\mathcal H}' = \tilde{g}\int_0^L \frac{dx}{2\pi} \cos(2\sqrt 2\phi_+).
    \label{H'_2sqrt2phi}
\end{align}
Just like in the spin chains, the coupling constant $\tilde{g}$ acquires the correction of $O(L^{-2})$,
\begin{align}
    \tilde{g} = g + g'\biggl(\frac{2\pi q}{L}\biggr)^2.
\end{align}
When the model is located at the quantum phase transition point (i.e. $g=0$), the perturbation to $\tilde{\mathcal{H}}_{\rm ladder}$ is written as
\begin{align}
    \tilde{\mathcal H}' &= g'\biggl(\frac{2\pi q}{L}\biggr)^2 \int_0^L \frac{dx}{2\pi} \cos(2\sqrt 2\phi_+).
\end{align}
Therefore, the polarization amplitude up to the first-order perturbative expansion $z^{(q)} \simeq \delta_1 z^{(q)}$ follows the scaling law just like Eq.~\eqref{zq_J1J2} for the fine-tuned $J_1$-$J_2$ XXZ spin chain,
\begin{align}
    z^{(q)} 
    &\simeq -\frac{{}_0\braket{0|\tilde{\mathcal H}'|e^{i2\sqrt 2 q\phi_+}}_0}{E_{2\sqrt 2}-E_0}
    \notag \\
    &= -\frac{q^2g'}{2x_{2\sqrt 2}} (\delta_{q,1}+\delta_{q,-1}) \biggl(\frac{2\pi}{L}\biggr)^{x_{2\sqrt 2}}.
    \label{zq_ladder}
\end{align}
There are two differences between Eqs.~\eqref{zq_J1J2} and \eqref{zq_ladder}.
First, the power is different.
This is obviously due to the difference of the interaction in $\mathcal H'$.
Second, the value of $q$ that makes $z^{(q)}$ nonzero is different.
$z^{(1)}$ is nonzero in the spin ladder but $z^{(1)}=0$ in the spin chain.
The value $q$ that makes $z^{(q)}$ nonzero depends on the number of $S=1/2$ spins in the unit cell~\cite{nakamura-todo}.

In analogy with the $S=1/2$ XXZ spin chain, the scaling law is changed for $g\not=0$ in the $L\to + \infty$ limit.
\begin{align}
    z^{(1)} = -\frac{g}{2x_{2\sqrt 2}} \biggl(\frac{2\pi}{L}\biggr)^{x_{2\sqrt 2}-2}.
    \label{zq_ladder_offcritical}
\end{align}
Note that $x_{2\sqrt 2}=2K$ is smaller than $2$ because $\cos(2\sqrt 2\phi_+)$ is relevant.
The divergence \eqref{zq_ladder_offcritical} of $z^{(1)}$ in the $L\to +\infty$ limit can be understood as a manifestation of the  nonzero lowest-energy excitation gap from the ground state in the $L\to + \infty$ limit.

The scaling law \eqref{zq_ladder_offcritical} is valid only when the relevant interaction \eqref{H'_2sqrt2phi} can be regarded as a perturbation.
This condition is met when $L$ is much smaller than the correlation length $v/\Delta_0$ with $\Delta_0$ being the lowest-energy excitation gap from the ground state.

Fortunately, the exact matrix element ${}_0\braket{0|e^{i2\sqrt 2\phi_+}|0}_0$ is available in the gapped phase of the sine-Gordon theory \eqref{H_ladder_eff}.
For $g>0$, it is given by~\cite{lukyanov_sinegordon}
\begin{widetext}
\begin{align}
    {}_0\braket{0|e^{i2\sqrt 2\phi_+(x)}|0}_0
    &= \biggl[\frac{a_0\Delta_0 \Gamma(\frac 1{2-2K})}{2v\Gamma(\frac{K}{2-2K})}
    \biggr]^{2K}\exp\biggl\{\int_0^\infty \frac{dt}{t}\biggl[-2K e^{-2t}+ \frac{\sinh^2(2Kt)}{2\sinh(K t)\sinh t \cosh((1-K)t)}\biggr]
    \biggr\}.
    \label{exp_gapped}
\end{align}
\end{widetext}
The gapped phase for $g<0$ has one-to-one correspondence to the one for $g>0$ via the shift $\phi_+ \to \phi_+ + \frac{\pi}{2\sqrt 2}$ which goes with the sign change of the vertex operator, $e^{i2\sqrt 2\phi_+} \to - e^{i2\sqrt 2\phi_+}$.
Therefore $z^{(q)}$ for $g\not=0$ converges to a finite value in the $L\to +\infty$ limit and
\begin{align}
    \lim_{L\to + \infty} z^{(q)} &\propto \operatorname{sgn}(g) \biggl(\frac{\Delta_0}{v} \biggr)^{x_{2\sqrt 2}}(\delta_{q,1}+\delta_{q,-1}),
    \label{zq_ladder_delta}
\end{align}
where $\operatorname{sgn}(g)$ denotes the sign of $g$.
Therefore, we obtain
\begin{align}
    \lim_{g\to 0}\lim_{L\to +\infty}z^{(q)} = \lim_{L\to + \infty}\lim_{g\to 0} z^{(q)} = 0.
    \label{zq_limit_ladder}
\end{align}
Therefore, the polarization amplitude $z^{(q)}$ in the $L\to + \infty$ limit is a continuous function of $g$ and vanishes at the quantum critical point smoothly.

\section{Universal jump of the polarization amplitude}\label{sec:APBC}

\subsection{Antiperiodic boundary condition}

We showed that the scaling law \eqref{zq_scaling} is universal in a sense that it is explained in terms of the low-energy effective field theory once the form of the perturbation to the fixed-point theory is identified.
Simultaneously, the scaling law turned out not to be uniquely determined by the fixed-point theory alone.
It depends on some details of the perturbation.
In this section, we discuss a property of the polarization amplitude determined fully by the fixed-point theory itself and independent of the perturbations.

Thus far we have imposed the periodic boundary condition \eqref{pbc_spin} on the spin.
Here instead, we consider the spin operator $\tilde{\bm S}_j$ satisfying the antiperiodic boundary condition,
\begin{align}
    \tilde S_{L+1}^\pm &= - \tilde S_{1}^\pm , \qquad \tilde S_{L+1}^z= \tilde S_1^z.
    \label{apbc_spin}
\end{align}
Let us focus on the simplest case, the $S=1/2$ XXZ spin chain.
Under the antiperiodic boundary condition,
its Hamiltonian is modified to be
\begin{align}
    \mathcal H_{\rm AP} &= J \sum_{j=1}^{L-1} (\tilde S_j^x\tilde S_{j+1}^x+\tilde S_j^y\tilde S_{j+1}^y + \Delta \tilde S_j^z \tilde S_{j+1}^z) 
    \notag \\
    &\qquad + J [-(\tilde S_L^x\tilde S_1^x + \tilde S_L^y \tilde S_1^y) +\Delta \tilde S_L^z \tilde S_1^z].
    \label{H_XXZ_A}
\end{align}
It is well known~\cite{fukui_twist, kitazawa_tbc, hirano_bp} that the $S=1/2$ XXZ spin chain under the antiperiodic boundary condition has doubly degenerate ground states in the $L\to + \infty$ limit.
Let us denote them as $\ket{\psi_{0,n}}$ and $n=\pm$.

One can see easily the double degeneracy by writing down the low-energy effective theory.
We take several steps
to reach the low-energy effective form of the Hamiltonian \eqref{H_XXZ_A}.
Let us first consider the following transformation:
\begin{align}
    \mathcal H_{\rm P}
    &:= U^{\frac 12}\mathcal{H}_{\rm AP} U^{-\frac 12}.
    \label{tildeH_XXZ_A}
\end{align}
$\mathcal{H}_{\rm P}$ is identical to the $S=1/2$ XXZ spin chain under the periodic boundary condition.
Let us relate $\tilde{\bm S}_j$ to the spin $\bm S_j$ satisfying the periodic boundary condition through
\begin{align}
    S_j^\pm &= e^{\pm i \frac{\pi}{L}j }\tilde S_j^\pm, \qquad S_j^z = \tilde S_j^z,
\end{align}
for $j=1,2, \cdots, L$.
The transformed spin operator indeed obeys the SU(2) commutation relation and the periodic boundary condition \eqref{pbc_spin}.
In addition, $\mathcal{H}_{\rm P}$ is simply given by
\begin{align}
    \mathcal H_{\rm P}
    &= J \sum_{j=1}^L (S_j^x  S_{j+1}^x + S_j^y S_{j+1}^y + \Delta  S_j^z S_{j+1}^z),
\end{align}
which becomes the TL liquid Hamiltonian \eqref{H_XXZ_fixedpoint} plus irrelevant perturbations $\mathcal H'$ at low energies.
Next, performing the inverse transformation, 
\begin{align}
    \mathcal H_{\rm AP} = U^{-1/2}\mathcal H_{\rm P} U^{1/2},
\end{align}
we obtain
\begin{align}
    \mathcal H_{\rm AP}
    &= \frac{v}{2\pi}\int_0^L dx \biggl[ K \biggl( \partial_x \theta  - \frac{\pi}{L} + \pi \delta(x-L) \biggr)^2 
    \notag \\
    &\qquad + \frac{1}{K}(\partial_x \phi)^2 \biggr]
    +\mathcal H'.
    \label{H_XXZ_A_eff}
\end{align}
The perturbation $U^{1/2}\mathcal H'U^{-1/2}$ is approximated as $\mathcal H'$ in Eq.~\eqref{H_XXZ_A_eff} similarly to the XXZ chain in the periodic boundary condition.

Thus, the vacuum and the highest-weight state corresponding to the vertex operator $e^{i2n\phi}$ of the model \eqref{H_XXZ_A_eff} with $\mathcal H'=0$ are given by
$\ket{0}_{q=-1/2}$ and $\ket{e^{i2n\phi}}_{q=-1/2}$, respectively, which are identified with~\cite{kitazawa_tbc}
\begin{align}
    \ket{0}_{-1/2} &= \ket{e^{-i\phi}}_0, \\
    \ket{e^{i2n\phi}}_{-1/2} &=  \ket{e^{i(2n-1)\phi}}_0.
\end{align}
One will be aware of a fact that $\ket{0}_{-1/2}$ and $\ket{e^{i2\phi}}_{-1/2}$ correspond to $\ket{e^{-i\phi}}_0$ and $\ket{e^{i\phi}}_0$, respectively, and are energetically degenerate thanks to a $\mathbb Z_2$ symmetry of the Hamiltonian $\mathcal H_{\rm P}$.
The symmetry is the invariance under the $\phi(\tau,x) \to - \phi(\tau,x)$ transformation.
Respecting the $\phi\to -\phi$ symmetry, we can reconstruct $\ket{e^{\pm i\phi}}_0$ as
\begin{align}
    \ket{\cos\phi}_0 &:= \frac{\ket{e^{i\phi}}_0 +\ket{e^{-i\phi}}_0}{\sqrt 2},
    \label{ket_cos} \\
    \ket{\sin\phi}_0
    &:= \frac{\ket{e^{i\phi}}_0 - \ket{e^{-i\phi}}_0}{\sqrt 2i}.
\end{align}

The condition of $\phi(\tau,x) \to -\phi(\tau,x)$ can be relaxed.
An operation $\phi(\tau,x) \to - \phi(\tau,L-x)$ plays the same role.
For example, the site-centered inversion $\mathcal I_s : \bm S_j \to \bm S_{L-j}$ for even $L$ acts on $\phi(\tau,x)$ as
\begin{align}
    \mathcal I_s\phi(\tau,x) \mathcal I_s^{-1} &= - \phi(\tau,L-x).
\end{align}
$\mathcal I_s\ket{e^{\pm i\phi}}_0 = \ket{e^{\mp i\phi}}_0$ holds true.

Under the $\phi\to-\phi$ symmetry, every eigenstate of the Hamiltonian can simultaneously be an eigenstate of the corresponding symmetry operator such as $\mathcal I_s$.
The doubly degenerate ground states respecting the site-centered inversion symmetry are
\begin{align}
    \ket{\psi_{0,+}} &= \ket{\cos\phi}_0,
    \label{psi_0+} \\
    \ket{\psi_{0,-}} &= \ket{\sin\phi}_0.
    \label{psi_0-}
\end{align}

Two kinds of polarization amplitude follow from the doubly degenerate ground states.
\begin{align}
    z_\pm^{(q)} := \braket{\psi_{0,\pm}|U^q|\psi_{0,\pm}}.
    \label{zpm_def}
\end{align}
$z_\pm^{(q)}$ are formalized in terms of the TL liquid.
\begin{align}
    z_\pm^{(q)}
    &= \frac 12 \biggl[
    {}_0\braket{e^{i\phi}|e^{i(2q+1)\phi}}_0 + {}_0 \braket{e^{-i\phi}|e^{i(2q-1)\phi}}_0
    \notag \\
    &\qquad \pm {}_0 \braket{e^{i\phi}|e^{i(2q-1)\phi}}_0
    \pm {}_0\braket{e^{-i\phi}|e^{i(2q+1)\phi}}_0
    \biggr].
    \label{zpm_cft}
\end{align}
The last two terms of Eq.~\eqref{zpm_cft} can be nonzero for $q\not=0$ at zero-th order of the perturbation.
In fact, we obtain
\begin{align}
    z_\pm^{(q)}
    &= \pm \frac 12 (\delta_{q,-1}+\delta_{q,1}),
    \label{zq_AP}
\end{align}
without requiring any perturbation to the fixed-point theory.
Moreover, the smallest positive $q$ that makes $z^{(q)}$ nonzero is $1$ instead of $2$ under the periodic boundary condition.

The relation \eqref{zq_AP} means that the polarization amplitude shows a discontinuity in the presence of the $\phi\to-\phi$ symmetry.
To clarify the claim, we modify the TL-liquid Hamiltonian $\mathcal H_{\rm P}$ without flux to
\begin{align}
    \mathcal H_{\rm P}
    &= \frac{v}{2\pi} \int_0^L dx \biggl( K(\partial_x \theta)^2 + \frac 1K(\partial_x \phi)^2 \biggr)
    \notag \\
    &\qquad + h \int_0^L \frac{dx}{2\pi} \cos(2\phi).
\end{align}
Introduction of $\cos(2\phi)$ corresponds to application of the staggered magnetic field $h$ coupled to $\sum_j (-1)^j S_j^z$ to the spin chain.
Up to the first order of $h$,
The eigenenergy $E_{0,\pm}$ of $\ket{\psi_{0,\pm}}_0$ is shifted by $h\int_0^L (dx/2\pi){}_0\braket{\psi_{0,\pm}|\cos(2\phi)|\psi_{0,\pm}}_0$, that is,
\begin{align}
    E_{0,\pm}
    &= \frac{2\pi}{L} \biggl(x_1 \pm \frac{h}{2}\biggl(\frac{2\pi}{L}\biggr)^{x_2-2}\biggr).
\end{align}
Therefore, $\ket{\psi_{0,+}}_0$ is the unique ground state for $h<0$ and $\ket{\psi_{0,-}}_0$ for $h>0$.
The polarization amplitude in the vicinity of the quantum critical point $h=0$ is well-defined except for $h=0$.
\begin{align}
    z^{(1)}
    &=\left\{
    \begin{array}{cc}
        z_+^{(1)}, & (h<0), \\
        & \\
        z_-^{(1)}, & (h>0).
    \end{array}
    \right.
\end{align}
It immediately follows that
\begin{align}
    \lim_{h\to -0} z^{(1)} &= \frac 12, \\
    \lim_{h\to +0} z^{(1)} &= - \frac 12.
\end{align}
The polarization amplitude $z^{(1)}$ depends on a way to eliminate the relevant interaction.
It is in sharp contrast to the spin chains and the spin ladder under the periodic boundary condition.
In the periodic boundary condition,
the polarization amplitude smoothly vanishes when going across the quantum critical point.
The universal value of jump,
\begin{align}
    \Delta z^{(1)} := \lim_{h\to -0} z^{(1)} - \lim_{h\to +0} z^{(1)} = 1
    \label{jump_z1}
\end{align}
reflects the $\mathbb Z_2$ symmetry at $h=0$.
As already pointed out in Ref.~\cite{nakamura_jump}, the universal jump \eqref{jump_z1} enables us to detect precisely the phase transition at $h=0$ pursuing $z^{(q)}$ as a function of $h$ under the antiperiodic boundary condition more clearly than doing under the periodic boundary condition.

\subsection{Physical origin of the jump}

Here, we focus on symmetrical origin of the universal jump \eqref{jump_z1} occurring in association with the LSM theorem.
Let us recall the fact that $\ket{\psi_0}$ is orthogonal to $U\ket{\psi_0}$.
This fact is crucial in the proof of the LSM theorem~\cite{lsm}.
Also, it is the orthogonality of $U\ket{\psi_0}$ and $\ket{\psi_0}$ to make the ground state under the antiperiodic boundary condition doubly degenerate.
However, the double degeneracy of the ground state does not immediately result in the universal jump.

Note that there are options of choice to construct the eigenstate of the Hamiltonian based on symmetry.
In Eqs.~\eqref{psi_0+} and \eqref{psi_0-}, we took the eigenstates $\ket{\psi_{0,\pm}}$ to be the simultaneous eigenstate of $\mathcal I_s$:
\begin{align}
    \mathcal I_s \ket{\psi_{0,\pm}} = \pm \ket{\psi_{0,\pm}}.
    \label{Is-on-psi_0_pm}
\end{align}
We can define $\ket{\psi_{0,\pm}}$ to be the simultaneous eigenstate of the symmetry operation of $T_1$.
Since $T_1\ket{e^{\pm i\phi}}_0 = \pm i \ket{e^{\pm i\phi}}_0$, we can define
\begin{align}
    \ket{\psi_{0,\pm}} = \ket{e^{\pm i\phi}}_0
    \label{psi_0_pm_T1}
\end{align}
to respect the one-site translation symmetry of $\mathcal H_{\rm P}$.

The choice \eqref{psi_0_pm_T1} of $\ket{\psi_{0,\pm}}$ makes $z^{(1)}_\pm$ continuous.
In fact, $z^{(1)}_\pm = 0$.
We can ask a question of what symmetry we should impose on the system to realize nonzero $z^{(1)}$ under the antiperiodic boundary condition.
Recall that $\ket{\psi_{0,\pm}}$ of Eqs.~\eqref{psi_0+} and \eqref{psi_0-} lead to $z^{(1)}_\pm \not=0$ because $\ket{\cos\phi}_0$ and $\ket{\sin\phi}_0$ are superpositions of $U^{1/2}\ket{\psi_0}$ and $U^{-1/2}\ket{\psi_0}$,
\begin{align}
    \ket{\cos\phi}_0 &= \frac{U^{\frac 12}+U^{-\frac 12}}{\sqrt 2}\ket{\psi_0}, \\
    \ket{\sin\phi}_0 &= \frac{U^{\frac 12}-U^{-\frac 12}}{\sqrt 2 i}\ket{\psi_0}.
\end{align}
This representation of $\ket{\cos\phi}_0$ and $\ket{\sin\phi}_0$ is consistent with a relation,
\begin{align}
    \mathcal I_s U \mathcal I_s^{-1} &= U^{-1}.
    \label{Is-U}
\end{align}
The essence of the relation \eqref{Is-U} is clarified when we represent the operator $U=\exp(2\pi i \mathcal P/L)$ using the naive polarization operator,
\begin{align}
    \mathcal P = \sum_{j=1}^L j S_j^z.
\end{align}
The relation \eqref{Is-U} is rephrased as
\begin{align}
    \mathcal I_s \mathcal P \mathcal I_s^{-1} = -\mathcal P + L \sum_{j=1}^L S_j^z.
    \label{Is-P}
\end{align}
Namely, when the ground state has zero total magnetization $\sum_{j=1}^L S_j^z=0$, the site-centered inversion operator $\mathcal I_s$ anticommutes with the parity operator $\mathcal P$.
This is in contrast to the one-site translation $T_1$ that anticommutes with $U$,
\begin{align}
    T_1UT_1^{-1} = -U,
    \label{T1-U}
\end{align}
rather than $\mathcal P$.
The one-site translation symmetry 
\begin{align}
    T_1 \mathcal H_{\rm AP} T_1^{-1} &= \mathcal H_{\rm AP},
    \label{T1-in-AP}
\end{align}
is obvious if we write it in terms of $\bm S_j$ as
\begin{align}
    \mathcal H_{\rm AP}
    &= \sum_{j=1}^L \biggl( \frac{e^{i\pi/L}}2 S_j^+S_{j+1}^- + \frac{e^{-i\pi/L}}2 S_j^-S_{j+1}^+ +\Delta S_j^z S_{j+1}^z \biggr).
\end{align}
On the other hand, the Hamiltonian $\mathcal H_{\rm AP}$ has the site-centered inversion symmetry $\mathcal I_s$ in the sense of the large gauge transformation:
\begin{align}
    \mathcal I_s \mathcal H_{\rm AP} \mathcal I_s^{-1}
    &= U \mathcal H_{\rm AP} U^{-1}.
    \label{Is-in-AP}
\end{align}
This difference of the symmetries in association with $U$ determines the fate of $z^{(1)}$.

The above argument around the relation of $\mathcal I_s$ to $z^{(1)}$ is generalized as follows.
Let $\ket{\psi_0}$ be the unique ground state of a generic Hamiltonian $\mathcal H_{\rm gen}$ of one-dimensional spin system with the chain length $L$: $\mathcal H_{\rm gen} \ket{\psi_0} = E_0 \ket{\psi_0}$.
Here we perform the following transformation,
\begin{align}
    \tilde{\mathcal H}_{\rm gen} &= U^{1/2}\mathcal H_{\rm gen} U^{-1/2}.
\end{align}
In addition, we assume
\begin{align}
    \lim_{L \to + \infty} \tilde{\mathcal H}_{\rm gen} &= \mathcal H_{\rm gen}.
    \label{H_gen_lim}
\end{align}
The ground state of $\tilde{\mathcal H}_{\rm gen}$ is obviously $\ket{\psi_{1/2}} = U^{1/2} \ket{\psi_0}$.
Let us assume the existence of an operator $\mathcal O_{\rm sym}$ that satisfies
\begin{align}
    \mathcal O_{\rm sym} \mathcal H_{\rm gen} \mathcal O_{\rm sym} &= \mathcal H_{\rm gen}, \\
    \mathcal O_{\rm sym} U^{1/2} \mathcal O_{\rm sym}^{-1} &= e^{i\Theta} U^{-1/2},
    \label{OUO-1}
\end{align}
with $\Theta\in \mathbb R$.
The latter relation leads to
\begin{align}
    \mathcal O_{\rm sym}\tilde{\mathcal H}_{\rm gen} \mathcal O_{\rm sym}^{-1} &= U^{-1} \tilde{\mathcal H}_{\rm gen} U.
\end{align}
The $\mathcal O_{\rm sym}$ commutes with $\tilde{\mathcal H}_{\rm gen}$ only for $L\to + \infty$. Then, $\ket{\psi_{-1/2}} = U^{-1/2}\ket{\psi_0}$ is also the ground state of $\tilde{\mathcal H}_{\rm gen}$.

In the $L\to + \infty$ limit, we can take every eigenstate of the Hamiltonian $\tilde{\mathcal H}_{\rm gen}$ as a simultaneous eigenstate of $\mathcal O_{\rm sym}$.
The doubly degenerate ground states are then given by
\begin{align}
    \ket{\psi_\pm} &= \frac{U^{1/2}\pm  e^{i\Theta}U^{-1/2}}{\sqrt 2}\ket{\psi_0}.
\end{align}
In fact, they satisfy
\begin{align}
    \mathcal O_{\rm sym} \ket{\psi_\pm} &= \pm \ket{\psi_\pm},
\end{align}
and
\begin{align}
    z_\pm^{(1)} &= \pm \frac 12 e^{i\Theta}.
    \label{zpm_Theta}
\end{align}

In short, the double degeneracy of the ground state of $\mathcal H_{\rm AP}$ results from the LSM theorem.
Another symmetry satisfying the relation \eqref{OUO-1} is required to visualize the ground state degeneracy as the jump of the polarization amplitude $z_\pm^{(1)}$.

To conclude this section, we point out that the $\mathbb Z_2$ value $z_\pm^{(1)} = \pm 1$ in the spin ladder system corresponds to the $\mathbb Z_2$ value of the Berry phase~\cite{hirano_bp, maruyama_bp, chepiga_BP_ladder}.
Imposing the antiperiodic boundary condition is equivalent to imposing a local twist~\cite{hirano_bp_J1J2, hirano_bp}.
The latter is directly related to the Berry phase.
Therefore, the universal jump as well as the Berry phase captures the same topological property of gapped phases.
It is an interesting feature of the polarization amplitude that it contains much information of the ground state deep in the gapped phases and also in the vicinity of the quantum critical point.

\section{Summary}\label{sec:summary}

Let us summarize the paper.
The scaling law \eqref{zq_scaling} of $z^{(q)}$
depends on the most relevant interaction in $\tilde{\mathcal H}'$ which perturbs the fixed-point theory of the large-gauge transformed system in the interesting way.
The scaling law for the three models discussed in Sec.~\ref{sec:XXZ}, \ref{sec:J1J2}, and \ref{sec:HS} are dealt with on equal footing.
The perturbation $\tilde{\mathcal H}'$ in those three models are special cases of the following interaction:
\begin{align}
    \tilde{\mathcal H}' = \int_0^L \frac{drdr'}{(2\pi)^2} \gamma_L(|r-r'|) [J_R^x(r)J_L^x(r') + J_R^y(r) J_L^y(r')],
    \label{H'_chain_comp}
\end{align}
with a distance-dependent coupling constant $\gamma_L(|r-r'|)$.
In Secs.~\ref{sec:XXZ} and \ref{sec:J1J2}, we considered a local interaction, that is,
\begin{align}
    \gamma_L(|r-r'|) = \gamma_L(0) \delta(r-r').
\end{align}
This case is further divided into two: the cases with $\lim_{L\to +\infty}\gamma_L(0)\not=0$ and $\lim_{L\to + \infty} \gamma_L(0)=0$.
The former is the XXZ chain in Sec.~\ref{sec:XXZ} and the latter is the fine-tuned $J_1$-$J_2$ XXZ chain in Sec.~\ref{sec:J1J2}.
In Sec.~\ref{sec:HS}, we considered the nonlocal interaction,
\begin{align}
    \gamma_L(|r-r'|) = \gamma_L(0)
\end{align}
with $\lim_{L\to + \infty} \gamma_L(0)=0$.
The power $\beta_q$ of the scaling law \eqref{zq_scaling} in those models is composed of two parts:
\begin{align}
    \beta_q = 4qK + \delta.
\end{align}
The first part $4qK$ results from the scaling dimension of $J_R^x(r)J_L^x(r')+J_R^y(r)J_L^y(r')$ and the second part $\delta$ is independent of $K$ but dependent on the range of perturbation and on the $L\to + \infty$ limit of the perturbation.

The fixed-point Hamiltonian of those spin chains are exposed only to the irrelevant (or marginal at most) perturbations.
In contrast, that of the spin ladder is exposed to the relevant interaction.
When the system is off the quantum critical point, the scaling law \eqref{zq_scaling} is valid when the relevant interaction can be seen as a perturbation, that is, when the chain length $L$ is short enough.
For large enough $L$, the polarization amplitude is insensitive to $L$ and determined by the lowest-energy excitation gap [Eq.~\eqref{zq_ladder_delta}].
When the parameter is fine-tuned so that the system is exactly on the quantum critical point, the polarization amplitude follows the sclaing law \eqref{zq_scaling} just like the fine-tuned $J_1$-$J_2$ XXZ chain.
These results are summarized in table~\ref{tab:scaling}.

Section~\ref{sec:APBC} was devoted to discussions on the property of the polarization amplitude determined by the fixed-point Hamiltonian alone.
Here, we imposed the antiperiodic boundary condition on the generic one-dimensional spin system.
We considered the symmetry whose generator $\mathcal O_{\rm sym}$ satisfies the relation \eqref{OUO-1}.
If we classify the eigenstates of the Hamiltonian by such a symmetry, the polarization amplitude in the gapless phase under the antiperiodic boundary condition turned out to be double valued [Eq.~\eqref{zpm_Theta}].
It immediately follows from Eq.~\eqref{zpm_Theta} that the polarization amplitude exhibits the universal jump at the quantum critical point as shown in Eq.~\eqref{jump_z1}.
We comment that the universal jump is a consequence of the LSM theorem.

\section*{Acknowledgments}

The authors thank Y. Fukusumi, Y. Horinouchi, R. Kobayashi, and M. Oshikawa for instructive discussions.

M.~N. acknowledges the Visiting Researcher's Program
of the Institute for Solid State Physics, the University of Tokyo.
This work is supported by JSPS KAKENHI Grant Number 17K05580.

\appendix

\bibliography{ref.bib}

\end{document}